\documentclass[aps,pre,reprint]{revtex4-1}

\usepackage{natbib}
\usepackage{amssymb,amsmath}

\usepackage{epsfig}
\usepackage{bm}
\usepackage{color}
\usepackage{graphicx}% Include figure files
\usepackage{hyperref}
\usepackage{enumerate}

\newcommand\beq{\begin{equation}}
\newcommand\eeq{\end{equation}}
\newcommand\beqa{\begin{eqnarray}}
\newcommand\eeqa{\end{eqnarray}}
\newcommand{\nn}{\nonumber\\}
\def\bal#1\eal{\begin{align}#1\end{align}}

\newcommand{\ma}{m_i}

\newcommand{\mb}{m_j}

\newcommand{\cc}{\mathbf{v}}

\newcommand{\kk}{\widehat{\bm{\sigma}}}

\newcommand{\ww}{\bm{\omega}}

\newcommand{\Ia}{I_i}

\newcommand{\da}{\sigma_i}

\newcommand{\db}{\sigma_j}

\newcommand{\ds}{\sigma}

\newcommand{\dab}{\sigma_{ij}}

\newcommand{\x}{\times}

\newcommand{\een}{\alpha_{ij}}

\newcommand{\esn}{\alpha}

\newcommand{\eet}{\beta_{ij}}

\newcommand{\est}{\beta}

\newcommand{\en}{\widetilde{\alpha}_{ij}}

\newcommand{\et}{\widetilde{\beta}_{ij}}

\newcommand{\mab}{m_{ij}}

\newcommand{\qab}{\kappa_{ij}}

\newcommand{\qa}{\kappa_{i}}

\newcommand{\qb}{\kappa_{j}}

\newcommand{\q}{\kappa}

\newcommand{\fa}{f_{i}}

\newcommand{\fb}{f_{j}}

\newcommand{\far}{f_{i}^{\text{rot}}}

\newcommand{\fbr}{f_{j}^{\text{rot}}}

\newcommand{\Tat}{T_{i}^{\text{tr}}}

\newcommand{\Tbt}{T_{j}^{\text{tr}}}

\newcommand{\Tt}{T^{\text{tr}}}

\newcommand{\Tar}{T_{i}^{\text{rot}}}

\newcommand{\Tbr}{T_{j}^{\text{rot}}}

\newcommand{\Tr}{T^{\text{rot}}}

\newcommand{\Qab}{J_{ij}}

\newcommand{\na}{n_i}

\newcommand{\nb}{n_j}

\newcommand{\zabt}{\xi_{ij}^{\text{tr}}}

\newcommand{\zt}{\xi^{\text{tr}}}

\newcommand{\zabr}{\xi_{ij}^{\text{rot}}}

\newcommand{\zr}{\xi^{\text{rot}}}

\newcommand{\al}{i}

\newcommand{\be}{j}

\newcommand{\tr}{{\text{tr}}}

\newcommand{\rot}{{\text{rot}}}

\newcommand{\wwwab}{\mathbf{w}_{ij}}

\newcommand{\chiab}{{\chi}_{ij}}

\begin{document}

\title{Energy nonequipartition in  gas mixtures  of inelastic rough hard spheres: The tracer limit}

\author{Francisco Vega Reyes$^{1}$}
\author{Antonio Lasanta$^{2}$}
\author{Andr\'es Santos$^{1}$}
\author{Vicente Garz\'o$^{1}$}
\affiliation{
$^1$Departamento de F\'{\i}sica and Instituto de Computaci\'on Cient\'{\i}fica Avanzada (ICCAEx), Universidad de Extremadura, 06006 Badajoz, Spain\\
$^2$Gregorio Mill\'an Institute of Fluid Dynamics,
Nanoscience and Industrial Mathematics,
Department of Materials Science and Engineering and Chemical Engineering,
Universidad Carlos III de Madrid, 28911 Legan\'es, Spain}

\begin{abstract}
The dynamical properties of a tracer or impurity particle immersed in a host gas of inelastic and rough hard spheres in the homogeneous cooling state is studied. Specifically, the breakdown of energy equipartition  as characterized by the tracer/host ratios of translational and rotational temperatures is analyzed by exploring a wide spectrum of values of the control parameters of the system (masses, moments of inertia, sizes, and coefficients of restitution). Three complementary approaches are considered. On the theoretical side, the Boltzmann and Boltzmann--Lorentz equations (both assuming the molecular chaos ansatz) are  solved by means of a multitemperature Maxwellian approximation for the velocity distribution functions. This allows us to obtain explicit analytical expressions for the temperature ratios. On the computational side, two different techniques are used. First, the  kinetic equations are numerically solved by the direct simulation Monte Carlo  (DSMC) method.  Second, molecular dynamics simulations for dilute gases are performed. Comparison between theory and simulations shows a general good agreement. This means that (i) the impact of the molecular chaos ansatz on the temperature ratios is not significant (except at high inelasticities and/or big impurities) and (ii) the simple Maxwellian approximation yields quite reliable predictions.
\end{abstract}

\date{\today}
\maketitle

\section{Introduction}
\label{sec1}
It is well known that granular gases are
intrinsically out of equilibrium \cite{BP04,RN08}. As a consequence, energy  is in general unevenly distributed among the
different degrees of freedom,
even in homogeneous and isotropic states. For instance, if the grains have both translational and
rotational degrees of freedom, the mean kinetic energy associated with each degree of
freedom is in general different
\cite{GS95,HZ97,ML98,LHMZ98,HHZ00,CLH02,GNB05,Z06,TWV11,S11a,VT12,KSG14,VSK14,VSK14b,VS15}.
Analogously, in a granular mixture,  each component may have a different mean
translational kinetic energy \cite{GD99b,WP02,FM02,DHGD02,BT02a,MG02b,BRM05}. Of
course, the study of the lack of energy equipartition becomes much more complex
when both rotational degrees of freedom and polydispersity are considered.

Taking into account the rotational degrees of freedom is in general more realistic than ignoring them,
since they tend to be relevant in experiments, even for systems of spherical
particles; i.e., frictional forces may in general give rise to a spinning
motion, upon collision. The simplest way to model this effect is to consider that
collisions between spheres $i$ and $j$ are characterized by two constant coefficients
of normal ($\alpha_{ij}$) and tangential ($\beta_{ij}$) restitution
\cite{JR85a,ZTPSH98,PZMZ02}. The  coefficient of normal restitution $0<\alpha_{ij}<1$
helps characterize the decrease in the magnitude of the normal component of the
relative velocity of the two colliding spheres. On the other hand, the coefficient of
{tangential} restitution $-1\leq\beta_{ij}\leq 1$ accounts for the change in the
tangential component of the relative velocity at contact.

To the best of our knowledge, apart from works  on a fixed particle immersed in a bath
of thermalized point particles \cite{VT04,CP08}, the only studies of the nonequipartion of energy in a multicomponent gas of inelastic rough hard spheres were carried out in Refs.\ \cite{SKG10,S11b}.
In Ref.\ \cite{SKG10}, starting from the
Bogoliubov--Born--Green--Kirkwood--Yvon (BBGKY) hierarchy \cite{BDS97},
the collisional energy production rates  associated with the
translational and rotational temperatures were derived with the help of two main
approximations: (i) the factorization of the (precollisional) two-body velocity
distribution function as the product of two one-body distributions (i.e.,
\emph{molecular chaos}); and (ii) the neglect of correlations between translational and rotational
velocities, the translational velocity distribution function being
approached in the form of a Maxwellian distribution. In the case of a binary mixture in
the so-called homogeneous cooling state (HCS) \cite{H83}, the three independent temperature ratios ($\Tt_1/\Tt_2$,  $\Tr_1/\Tr_2$, and $\Tr_2/\Tt_2$) were  determined as functions of the three
coefficients of normal restitution ($\esn_{11}$, $\esn_{12}$, $\esn_{22}$), the three coefficients of tangential
restitution ($\est_{11}$, $\est_{12}$, $\est_{22}$), the two reduced moments of inertia ($\q_1$ and $\q_2$), the mass ratio ($m_1/m_2$), the
size ratio ($\ds_1/\ds_2$), and the two packing fractions  $\phi_i =\frac{\pi}{6}n_i\sigma_i^3$ ($i=1,2$), $n_i$ being the number density of component $i$.

It is obviously interesting to assess the degree of accuracy of the temperature ratios derived in Ref.\ \cite{SKG10} from approximations (i)
and (ii). Therefore, the main goal of this paper is to comparatively
study this approximate theoretical solution together with the exact numerical solution of the
corresponding kinetic equations  obtained via the direct simulation Monte Carlo
method (DSMC), as an assessment of approach (ii), and with molecular
dynamics (MD), as an additional and independent assessment of approaches (i) and (ii). On the other hand, since the parameter space  in a general binary mixture of rough spheres is $12$-dimensional, here we consider the simpler case where one of the components (say $i=1$) is present in tracer concentration (i.e., $n_1/n_2\to 0$). This limit case is equivalent to the problem of an impurity or intruder immersed in a granular gas of rough spheres (component $i=2$). This implies that the state of the gas bath is not affected by the presence of the tracer particles, so that its distribution function $f_2$ obeys, under the molecular chaos ansatz (i),  the Boltzmann equation for a one-component granular gas. Moreover, collisions among tracer particles are neglected. As a consequence, the velocity distribution $f_1$ of the tracer component obeys a Boltzmann--Lorentz kinetic equation. In this limit case, assuming that the state of the excess (or host) component is known \cite{ML98,LHMZ98}, the relevant temperature ratios are $\Tt_1/\Tt_2$ and  $\Tr_1/\Tr_2$, and the independent control parameters (apart from those associated with $f_2$, namely  $\phi_2$, $\esn_{22}$, $\est_{22}$, and $\q_2$) are $\esn_{12}$, $\est_{12}$, $\q_1$, $m_1/m_2$, and $\ds_1/\ds_2$.

In this paper we take the tracer limit from the more general results of Ref.\ \cite{SKG10} and carry out DSMC \cite{B94} and MD \cite{L91b} simulations to gauge the reliability of the theoretical predictions. Since the DSMC method is equivalent to a numerical solution of the Boltzmann equation, comparison between theory and DSMC results assess the assumption (ii)  described above. On the other hand, MD simulations are free from the molecular chaos hypothesis and thus they measure the impact of the assumption (i) on the temperature ratios. Some preliminary results were recently reported in Ref.\ \cite{VLSG17}.

The paper outline is as follows. For completeness, the kinetic theory for multicomponent granular gases is briefly summarized in Sec.\ \ref{sec2}. Then, the tracer limit is explicitly worked out in Sec.\ \ref{sec3}, where it is shown that the temperature ratios are given in terms of the solution of a quartic equation. Section \ref{sec4} is the core of the paper, as it deals with the comparison between the theoretical predictions and the computer simulations (both DSMC and MD) for $10$ representative classes of systems. The main conclusions of the work are presented in the concluding Sec.\ \ref{sec5}.

\section{Energy production rates in  general granular mixtures}
\label{sec2}

Let us consider a granular gas of inelastic rough hard spheres with an indefinite
number of species, where  by species we refer to a set of identical
particles. Particles of component $i$ have a mass $\ma$, a diameter $\da$, and a
moment of inertia $\Ia$. We use the following definition of dimensionless  moment of inertia  for species $i$,
\beq
\qa\equiv\frac{4\Ia}{\ma\da^2},
\label{kappa}
\eeq
which may range from $\qa=0$ (mass concentrated on the center) to $\qa=\frac{2}{3}$
(mass concentrated on the surface). A relevant case is $\qa=\frac{2}{5}$, for which the mass is uniformly distributed.

As advanced in Sec.\ \ref{sec1},  collisions between particles of sets $i$
and $j$ will be characterized with constant coefficients of normal and tangential restitutions $\een$ and $\eet$, respectively. They are defined by the collision rules
\beq
\kk\cdot \wwwab'=-\een \kk\cdot \wwwab,\quad \kk\x \wwwab'=-\eet \kk\x \wwwab,
\label{restitution}
\eeq
where $\wwwab$ and $\wwwab'$ are the pre- and post-collisional relative velocities of the points at contact and $\kk$ is the unit vector joining the centers of the two colliding spheres.
While the coefficient $\een$ ranges from $\een=0$ (perfectly inelastic particles) to $\een=1$ (perfectly elastic particles),  the coefficient $\eet$ runs from $\eet=-1$ (perfectly smooth particles) to $\eet=1$ (perfectly rough particles). Except if $\een=1$ and $|\eet|=1$, kinetic energy is dissipated upon a collision $ij$.

The relevant dynamic quantity is the velocity distribution function $f_i(\mathbf{v},\ww;t)$ of each component, where we have particularized to homogeneous states. Here, $\cc$ and $\ww$ denote the translational and rotational velocities, respectively. The number density and the translational and rotational velocities of component $i$ are defined, respectively, as
\begin{subequations}
\label{III.3}
\beq
 \na=\int d\cc \int d\ww\,  \fa(\cc,\ww),
\eeq
\beq
 \Tat=\frac{\ma}{3\na}\int d\cc \int d\ww\, v^2 \fa(\cc,\ww),
\eeq
\beq
 \Tar=\frac{\Ia}{3\na}\int d\cc \int d\ww\, \omega^2 \fa(\cc,\ww).
\eeq
\end{subequations}
The \emph{global} temperature is
\beq
T=\sum_\al\frac{\na}{2n}\left(\Tat+\Tar\right),
\label{III.13}
\eeq
where $n=\sum_\al \na$ is the total number density.

Assuming molecular chaos, the one-particle velocity distribution functions $\{f_i\}$ obey a closed set of coupled Boltzmann--Enskog equations \cite{RN08,SKG10},
\beq
\partial_t \fa(\cc_1,\ww_1;t)=\sum_\be \Qab[\cc_1,\ww_1;t|\fa,\fb],
\label{2}
\eeq
where
\bal
\Qab&[\cc_1,\ww_1;t|\fa,\fb]=\chi_{ij}\dab^2\int d\cc_2\int d\ww_2\int d\kk\,\nn &\times\Theta(\cc_{12}\cdot\kk)(\cc_{12}\cdot\kk)\Bigg[\frac{1}{\een^2\eet^2}\fa(\cc_1'',\ww_1'';t)\fb(\cc_2'',\ww_2'';t)\nn
&-
\fa(\cc_1,\ww_1;t)\fb(\cc_2,\ww_2;t)\Bigg]
\label{III.2}
\eal
is the collision operator. Here, $\chi_{ij}$ is the contact value of the spatial pair correlation function, $\dab\equiv (\da+\db)/2$, $\cc_{12}=\cc_1-\cc_2$ is the relative translational velocity, and the double primes denote precollisional velocities.
Taking moments in Eq.\ \eqref{2}, one can easily get the evolution equations for the partial temperatures as
\beq
\label{evol_T}
\partial_t \Tat=-\zt_i \Tat,\quad \partial_t \Tar=- \zr_i \Tar,
\eeq
with
\beq
\zt_i=\sum_j \zabt ,\quad \zr_i=\sum_j \zabr,
\eeq
where
\begin{subequations}
\label{54}
\beq
\zabt\equiv -\frac{\ma}{3\na\Tat}\int d\cc\int d\ww\, v^2 \Qab[\cc,\ww;t|\fa,\fb],
\label{54a}
\eeq
\beq
\zabr\equiv -\frac{\Ia}{3\na\Tar}\int d\cc\int d\ww\, \omega^2 \Qab[\cc,\ww;t|\fa,\fb]
\label{54b}
\eeq
\end{subequations}
are energy production rates. They are in general complex functionals of the distribution functions $\fa$ and $\fb$, so that the set of Eqs.\ \eqref{evol_T} is not closed.
The evolution equation for the global temperature is
\beq
\label{evol_TT}
\partial_t T=-\zeta T,
\eeq
where
\beq
\zeta=\sum_\al \frac{\na}{2nT}
\left(\Tat\xi^\tr_\al+\Tar\xi^\rot_\al\right)
\label{110}
\eeq
is the cooling rate. In contrast to the energy production rates $\zabt$ and $\zabr$, the cooling rate $\zeta$ is positive definite, i.e., collisions produce a decrease of the total temperature $T$ unless $\een=1$ and $\eet=\pm 1$ for \emph{all} pairs $\al\be$.

Even though $T$ monotonically decreases with time, it is expected
that, after a certain transient stage, a scaling regime is reached
where all the time dependence occurs through the total temperature
$T$. This is the so-called HCS \cite{H83}, which implies $\partial_t
( \Tat/T )=\partial_t ( \Tar/T )=0$, so that
\beq
\label{HCS}
\zt_i=\zr_i=\zeta
\eeq
for all components.

To express the production rates $\zabt$ and $\zabr$ in terms of the partial temperatures $\Tat$, $\Tar$, $\Tbt$, and $\Tbr$, we assume that the production rates can be \emph{estimated} by the replacements
\begin{subequations}
\label{IV.1}
\beq
\fa(\cc,\ww)\to  \left(\frac{\ma}{2\pi\Tat}\right)^{3/2}\exp\left(-\frac{\ma v^2}{2\Tat}\right)
\far(\ww),
\eeq
\beq
\fb(\cc,\ww)\to  \left(\frac{\mb}{2\pi\Tbt}\right)^{3/2}\exp\left(-\frac{\mb v^2}{2\Tbt}\right)
\fbr(\ww)
\eeq
\end{subequations}
in Eqs.\ \eqref{54}. Here,
\beq
\far(\ww)=\int d\cc\,\fa(\cc,\ww), \quad \fbr(\ww)=\int d\cc\,\fb(\cc,\ww)
\label{III.10}
\eeq
are marginal distributions that do not need to be known. The hypothesis behind  Eq.\ \eqref{IV.1} is two-fold. First, the statistical correlations between the translational and rotational velocities are ignored. Second, the translational marginal  distribution function is approximated by a Maxwellian.
It is important to stress that we are not making the strong
claim that $\fa$ and $\fb$ are well approximated by Eqs.\  \eqref{IV.1} in the
Boltzmann--Enskog equation \cite{BPKZ07,VSK14,VSK14b}, just that the production rates
can be approximately computed by performing those replacements.

By inserting Eqs.\ \eqref{IV.1} into Eqs.\ \eqref{54}, and after some algebra, one achieves the explicit expressions \cite{SKG10,S11b}
\begin{widetext}
\begin{subequations}
\label{55}
\beq
\zabt=\frac{4\nu_{\al\be}}{3\ma\Tat}
\left[2\left({\en+\et}\right){\Tat}-
\left({\en^2+\et^2}\right)\left(\frac{\Tat}{\ma}+\frac{\Tbt}{\mb}\right)
-{\et^2}\left(\frac{\Tar}{\ma\qa}+\frac{\Tbr}{\mb\qb}\right)
\right],
\label{55a}
\eeq
\beq
\zabr=\frac{4\nu_{\al\be}\et}{3\ma\qa\Tar}\left[2{\Tar}-{\et}\left(\frac{\Tat}{\ma}+\frac{\Tbt}{\mb}+\frac{\Tar}{\ma\qa}+\frac{\Tbr}{\mb\qb}\right)
\right],
\label{56}
\eeq
\beq
\zeta=\sum_{\al,\be} \frac{\na\mab\nu_{\al\be}}{3nT}\left[(1-\een^2)\left(\frac{\Tat}{\ma}+\frac{\Tbt}{\mb}\right)+\frac{\qab}{1+\qab}
(1-\eet^2)
\left(\frac{\Tat}{\ma}+\frac{\Tbt}{\mb}+\frac{\Tar}{\ma\qa}+\frac{\Tbr}{\mb\qb}\right)\right],
\label{114}
\eeq
\end{subequations}
\end{widetext}
where we have introduced the effective collision frequencies
\beq
\nu_{\al\be}\equiv\sqrt{2\pi}\chiab\nb{\dab^2}\sqrt{\frac{\Tat}{\ma}+\frac{\Tbt}{\mb}}
\label{56b}
\eeq
and
\begin{subequations}
\label{20}
\beq
\label{20a}
\en\equiv\mab\left(1+\een\right),\quad\et\equiv\frac{\mab\qab}{1+\qab}\left(1+\eet\right),
\eeq
\beq
\label{20b}
\mab\equiv \frac{\ma\mb}{\ma+\mb},\quad \qab\equiv \qa\qb\frac{\ma+\mb}{\qa\ma+\qb\mb}.
\eeq
\end{subequations}

In summary, in an $N_c$-component mixture, Eqs.\ \eqref{55}  allows one to obtain from Eq.\ \eqref{HCS} a closed set of $2N_c-1$  coupled algebraic equations for the $2N_c-1$ independent temperature ratios, which in general must be solved numerically.

\section{Tracer limit}
\label{sec3}

We now consider the special case of a system with two species where one of the components ($i=1$) is present in tracer concentration (i.e., $n_1/n_2\to 0$). In this limit, the Boltzmann Eqs.\ \eqref{2} decouple into a single Boltzmann equation for the host component ($i=2$) and a Boltzmann--Lorentz equation for the tracer component, namely
\begin{subequations}
\label{BL}
\beq
\partial_t f_2(\cc,\ww;t)=J_{22}[\cc,\ww;t|f_2,f_2],
\eeq
\beq
\partial_t f_1(\cc,\ww;t)=J_{12}[\cc,\ww;t|f_1,f_2].
\eeq
\end{subequations}
As a consequence of the limit $n_1/n_2\to 0$, the global temperature, \eqref{III.13}, and the cooling rate, \eqref{110}, reduce to
\beq
T=\frac{\Tt_2+\Tr_2}{2},\quad \zeta=\frac{\zt_2 \Tt_2+\zr_2 \Tr_2}{\Tt_2+\Tr_2}.
\eeq

\subsection{Host component}

In the Maxwellian approximation, Eq.\ \eqref{IV.1}, the production rates associated with the host component \cite{GS95,LHMZ98} can easily be obtained by setting $i=j=2$ in Eqs.\ \eqref{55}. Their expressions are
\begin{subequations}
\bal
\zt_{2}=&\zt_{22}=\frac{2\nu_{22}}{3}\Bigg[1-\esn_{22}^2
+\frac{2\q_{2}\left(1+\est_{22}\right)}{(1+\q_{2})^2}\left(1-\frac{\Tr_2}{\Tt_2}\right)\nn
&+\frac{\q_2\left(1-\est_{22}^2\right)}{(1+\q_2)^2}\left(\q_2+\frac{\Tr_2}{\Tt_2}\right)
\Bigg],
\label{61}
\eal
\bal
\zr_{2}=&\zr_{22}=\frac{2\nu_{22}}{3}\frac{1+\est_{22}}{(1+\q_2)^2}\frac{\Tt_2}{\Tr_2}\left[(1-\est_{22})\left(\q_2+\frac{\Tr_{22}}{\Tt_{22}}\right)
\right.\nn
&\left.-2\q_2\left(1-\frac{\Tr_2}{\Tt_2}\right)\right],
\label{62}
\eal
\beq
\zeta=\frac{2\nu_{22}}{3}\frac{\Tt_2}{\Tt_2+\Tr_2}\left[1-\esn_{22}^2+\frac{1-\est_{22}^2}{1+\q_2}\left(\q_2+\frac{\Tr_{2}}{\Tt_{2}}\right)\right].
\label{63}
\eeq
\end{subequations}

In the HCS, the condition $\zt_{2}=\zr_2$ yields the following quadratic equation for the temperature ratio $\Tr_2/\Tt_2$:
\beq
\frac{\Tr_2}{\Tt_2}-\frac{\Tt_2}{\Tr_2}=2\gamma_2,
\label{79}
\eeq
where
\beq
\label{gamma2}
\gamma_2\equiv \frac{(1+\kappa_2)^2}{2\kappa_2(1+\beta_{22})^2}\left[1-\alpha_{22}^2-\frac{1-\kappa_2}{1+\kappa_2}(1-\beta_{22}^2)\right].
\eeq
Its physical root is
\beq
\label{Tr2}
\frac{\Tr_2}{\Tt_2}=\sqrt{1+\gamma_2^2}+\gamma_2.
\eeq

It is interesting to note that the parameter $\gamma_2$ comprises completely the dependence of the temperature ratio on the set of parameters $\q_2$, $\alpha_{22}$, and $\beta_{22}$ in the Maxwellian approximation. The sign of that parameter results from the competition between two terms proportional to $1-\alpha_{22}^2$ and $1-\beta_{22}^2$, respectively. {}From Eq.\ \eqref{restitution} we observe that $1-\alpha_{22}^2=1-(\kk\cdot \mathbf{w}')^2/(\kk\cdot \mathbf{w})^2$ measures the relative decrease in the magnitude of the normal component of the relative velocity after a collision. Likewise,  $1-\beta_{22}^2=1-(\kk\x \mathbf{w}')^2/(\kk\x \mathbf{w})^2$ measures a similar relative decrease but in the case of the tangential component. Thus, $\gamma_2>0$ if the relative decrease of the normal component is larger than that of the tangential component [the latter being multiplied by $(1-\q_2)/(1+\q_2)$]; otherwise, $\gamma_2<0$. In the former case ($\gamma_2>0$), the mean rotational energy per particle is larger than the mean translational energy per particle, i.e., ${\Tr_2}/{\Tt_2}>1$, whereas  ${\Tr_2}/{\Tt_2}<1$ if $\gamma_2<0$. Equipartition of energy (${\Tr_2}/{\Tt_2}=1$) occurs if $\gamma_2=0$, implying a balance (in the sense described above) between the relative decrease of the magnitudes of the tangential and normal components of the relative velocity.

\subsection{Tracer component}
\label{sec3.B}

In the case of the tracer component $i=1$, one has $\zt_1=\zt_{12}$ and $\zr_1=\zr_{12}$, where $\zt_{12}$ and $\zr_{12}$, in the Maxwellian approximation, Eq.\ \eqref{IV.1}, can be obtained from Eqs.\ \eqref{55a} and \eqref{56}, respectively. In the HCS, Eq.\ \eqref{HCS} for $i=1$ yields
\begin{subequations}
\label{55bis}
\bal
\zeta^*=&s\frac{\sqrt{1+X}}{X}A\left[2(1+B) X-A(1+B^2)(1+X)\right.\nn
&\left.-AB^2r\left(1+Y\right)
\right],
\label{55.1}
\eal
\bal
\zeta^*=&s\frac{\sqrt{1+X}AB}{\kappa_1^2 r Y}\left[2\kappa_1 r Y-AB(1+X)\right.\nn
&\left.-ABr\left(1+Y\right)
\right].
\label{56.1}
\eal
\end{subequations}
Here, to simplify the notation, we have introduced the dimensionless quantities
\begin{subequations}
\beq
\label{XY}
X\equiv \frac{T_1^\tr}{T_2^\tr}\frac{m_2}{m_1},\quad Y\equiv \frac{T_1^\rot}{T_2^\rot}\frac{m_2\kappa_2}{m_1\kappa_1},
\eeq
\beq
\label{rs}
\zeta^*\equiv\frac{\zeta}{\nu_{22}},\quad r\equiv \frac{T_2^\rot}{\kappa_2T_2^\tr},\quad
 s \equiv \frac{2\sqrt{2}}{3}\frac{\chi_{12}\sigma_{12}^2}{\chi_{22}\sigma_2^2},
 \eeq
\beq
\label{AA}
A\equiv\frac{\widetilde{\alpha}_{12}}{m_1}=\frac{m_2}{m_1+m_2}(1+\alpha_{12}),\eeq
\beq
\label{BB}
B\equiv\frac{\widetilde{\beta}_{12}}{\widetilde{\alpha}_{12}}=\frac{\kappa_{12}}{1+\kappa_{12}}\frac{1+\beta_{12}}{1+\alpha_{12}}.
\eeq
\end{subequations}
The quantities $\zeta^*$, $r$, $s$, $A$, and $B$ are known in terms of the parameters of the problem, so that Eqs.\ \eqref{55bis} make a closed set of two nonlinear coupled equations for the unknowns $X$ and $Y$. To solve them, it is convenient to make the change of variable $X\to x^2-1$.  Equation \eqref{56.1} allows one to express the quantity $Y$ in terms of $x$. Next, insertion of $Y(x)$ into Eq.\ \eqref{55.1} yields a quartic equation for $x$, whose physical solution can easily be obtained.
Note that $X$ and $Y$ depend on the nine parameters of the problem only through the five quantities $\zeta^*$, $r$, $s$, $A$, and $B$.

\begin{table}
\caption{\label{tablelimit}Asymptotic behavior of the temperature ratios ${T_1^\tr}/{T_2^\tr}$ and ${T_1^\rot}/{T_2^\rot}$ in the limit $m_1/m_2\gg 1$, depending on the sign of  $\gamma_{12}$ [see Eq.\ \eqref{g12}]. The coefficients $\gamma_2$, $C_{\text{I}}$, $K_{\text{I}}$, $C_{\text{II}}$, $K_{\text{II}}$, and  $K_{\text{III}}$ are given by Eqs.\ \eqref{gamma2}, \eqref{CI}, \eqref{KI}, \eqref{CII}, \eqref{KII}, and \eqref{2.6}, respectively.}
\begin{ruledtabular}
\begin{tabular}{ccc}
$\gamma_{12}$&
${T_1^\tr}/{T_2^\tr}$&${T_1^\rot}/{T_2^\rot}$\\
\hline
$>0$&$\displaystyle{C_{\text{I}}^2\left(\frac{m_1}{m_2}\right)^3
\left(\frac{\sigma_{2}}{\sigma_{12}}\right)^4}$&$\displaystyle{\frac{K_{\text{I}}C_{\text{I}}^2}{\sqrt{1+\gamma_2^2}+\gamma_2}\left(\frac{m_1}{m_2}\right)^4
\left(\frac{\sigma_{2}}{\sigma_{12}}\right)^4}$\\
$=0$&$\displaystyle{C_{\text{I}}^2\left(\frac{m_1}{m_2}\right)^3
\left(\frac{\sigma_{2}}{\sigma_{12}}\right)^4}$&$\displaystyle{\frac{K_{\text{III}}C_{\text{I}}^2}{\sqrt{1+\gamma_2^2}+\gamma_2}\left(\frac{m_1}{m_2}\right)^3
\left(\frac{\sigma_{2}}{\sigma_{12}}\right)^4}$\\
$<0$&$\displaystyle{C_{\text{II}}^2\left(\frac{m_1}{m_2}\right)^3
\left(\frac{\sigma_{2}}{\sigma_{12}}\right)^4}$&$\displaystyle{\frac{K_{\text{II}}C_{\text{II}}^2}{\sqrt{1+\gamma_2^2}+\gamma_2}\left(\frac{m_1}{m_2}\right)^2
\left(\frac{\sigma_{2}}{\sigma_{12}}\right)^4}$\\
\end{tabular}
\end{ruledtabular}
\end{table}

\begin{table*}[htbp]
\caption{\label{cases}Parameters of the different systems considered in this work.}
\begin{ruledtabular}
\begin{tabular}{ccccccccc}
Label&$m_1/m_2$&$\sigma_1/\sigma_2$&$\kappa_2$&$\kappa_1$&$\alpha_{22}$&$\alpha_{12}$&$\beta_{22}$&$\beta_{12}$\\
\hline
A&$(\sigma_1/\sigma_2)^3$&$0.25$, \ldots, $2$&$\frac{2}{5}$&$\frac{2}{5}$&$0.9$&$0.9$&$-0.5$&$-0.5$\\
B&$(\sigma_1/\sigma_2)^3$&$0.25$, \ldots, $2$&$\frac{2}{5}$&$\frac{2}{5}$&$0.7$&$0.7$&$-0.5$&$-0.5$\\
C&$(\sigma_1/\sigma_2)^3$&$0.25$, \ldots, $2$&$\frac{2}{5}$&$\frac{2}{5}$&$0.9$&$0.9$&$0$&$0$\\
D&$(\sigma_1/\sigma_2)^3$&$0.25$, \ldots, $2$&$\frac{2}{5}$&$\frac{2}{5}$&$0.7$&$0.7$&$0$&$0$\\
E&$(\sigma_1/\sigma_2)^3$&$0.25$, \ldots, $2$&$\frac{2}{5}$&$\frac{2}{5}$&$0.9$&$0.9$&$0.5$&$0.5$\\
F&$(\sigma_1/\sigma_2)^3$&$0.25$, \ldots, $2$&$\frac{2}{5}$&$\frac{2}{5}$&$0.7$&$0.7$&$0.5$&$0.5$\\
G&$(\sigma_1/\sigma_2)^3$&$0.25$, \ldots, $2$&$0.1$&$0.1$&$0.2$&$0.2$&$1$&$1$\\
H&$1$&1&$\frac{2}{5}$&$\frac{2}{5}$&$0.9$&$0.5$, \ldots, $0.99$&$0$&$0$\\
I&$1$&1&$\frac{2}{5}$&$\frac{2}{5}$&$0.9$&0.9&$0$&$-0.99$, \ldots, $0.99$\\
J&$1$&1&$\frac{2}{5}$&$0.01$, \ldots, $\frac{2}{3}$&$0.9$&$0.9$&$0$&$0$\\
\end{tabular}
\end{ruledtabular}
\end{table*}

For completeness, the special limit $m_1\gg m_2$ (heavy impurity) is analyzed  in the Appendix. Table \ref{tablelimit} summarizes the asymptotic behaviors of the two tracer/host temperature ratios in that limit. Note that the  ratios depend on the sign of the quantity
\beq
\label{g12}
\gamma_{12}\equiv 1+\alpha_{12}-\frac{\kappa_2(1-\kappa_1)}{\kappa_1(1+\kappa_2)}(1+\beta_{12}).
\eeq
In all the cases, the tracer/host temperature ratios diverge in the limit $m_1/m_2\to\infty$ following power laws.

\section{Comparison with computer simulations}
\label{sec4}

In this section, the theoretical predictions derived in Sec.\ \ref{sec3} are compared
against computer simulations by the DSMC \cite{B94} and MD \cite{AW59,L91b} methods. The
first method obtains an exact numerical solution of the Boltzmann and
Boltzmann--Lorentz equations, and thus the assumption of molecular chaos is built
in. In contrast, the MD method avoids this bias since it is merely a solution of Newton's equations of motion and therefore is free from that assumption. In the DSMC simulations we have taken $\chi_{12}=\chi_{22}=1$, which corresponds to the dilute limit  $\phi_2\to 0$. On the other hand, in the MD simulations the value of the packing fraction needs to be finite and we have taken the values $\phi_2=\frac{\pi}{600}\simeq 0.005$ and $\phi_2=\frac{\pi}{300}\simeq 0.010$, which are small enough to justify the approximations $\chi_{12}\simeq\chi_{22}\simeq 1$.
Even in that case, there are eight independent control parameters of the mixture. Because of that, we have restricted ourselves to the 10 series of systems with parameters displayed in Table \ref{cases}.
In the series A--G, the tracer and host particles are assumed to be made of the same material, so that they share the values of the coefficients of restitution, the reduced moment of inertia, and the particle mass density. Thus the only independent parameter is the size ratio $\sigma_1/\sigma_2$ and, consequently, the mass ratio is $m_1/m_2=(\sigma_1/\sigma_2)^3$. While in cases A--F the mass distribution is uniform (so that $\q_1=\q_2=\frac{2}{5}$) and the particles are moderately inelastic, in case G the mass distribution is more concentrated near the center ($\q_1=\q_2=0.1$) and the particles are very inelastic and completely rough.
In the series H--J the tracer particles have the same size and mass as the host particles but they differ in the coefficient of normal  (case H) or tangential (case I) restitution, or in the reduced moment of inertia (case J). In the latter case the tracer and host particles are ``externally'' identical but they differ in their internal mass distribution.

\begin{figure}
\includegraphics[width=\columnwidth]{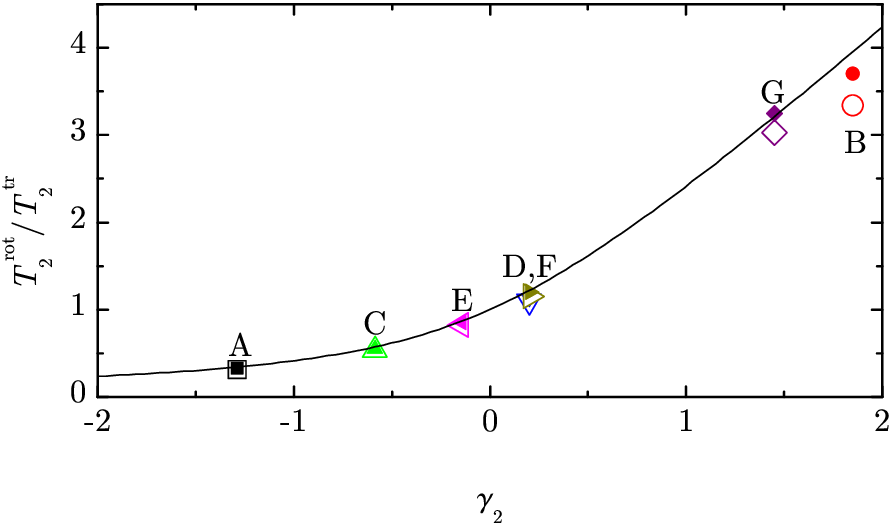}
\caption{Plot of the temperature ratio $T_2^\rot/T_2^\tr$ of the host gas versus the parameter $\gamma_2$ defined in Eq.\ \eqref{gamma2}.  The line is the theoretical prediction given by Eq.\ \eqref{Tr2}, while the filled and open symbols are DSMC and MD results, respectively. The symbols correspond to the cases A (squares), B (circles), C (up triangles), D (down triangles), E (left triangles), F (right triangles), and G (diamonds). The error bars in the simulation data are smaller than the size of the symbols.}
\label{fig:Tr2}
\end{figure}
 \begin{figure*}
\includegraphics[width=1.8\columnwidth]{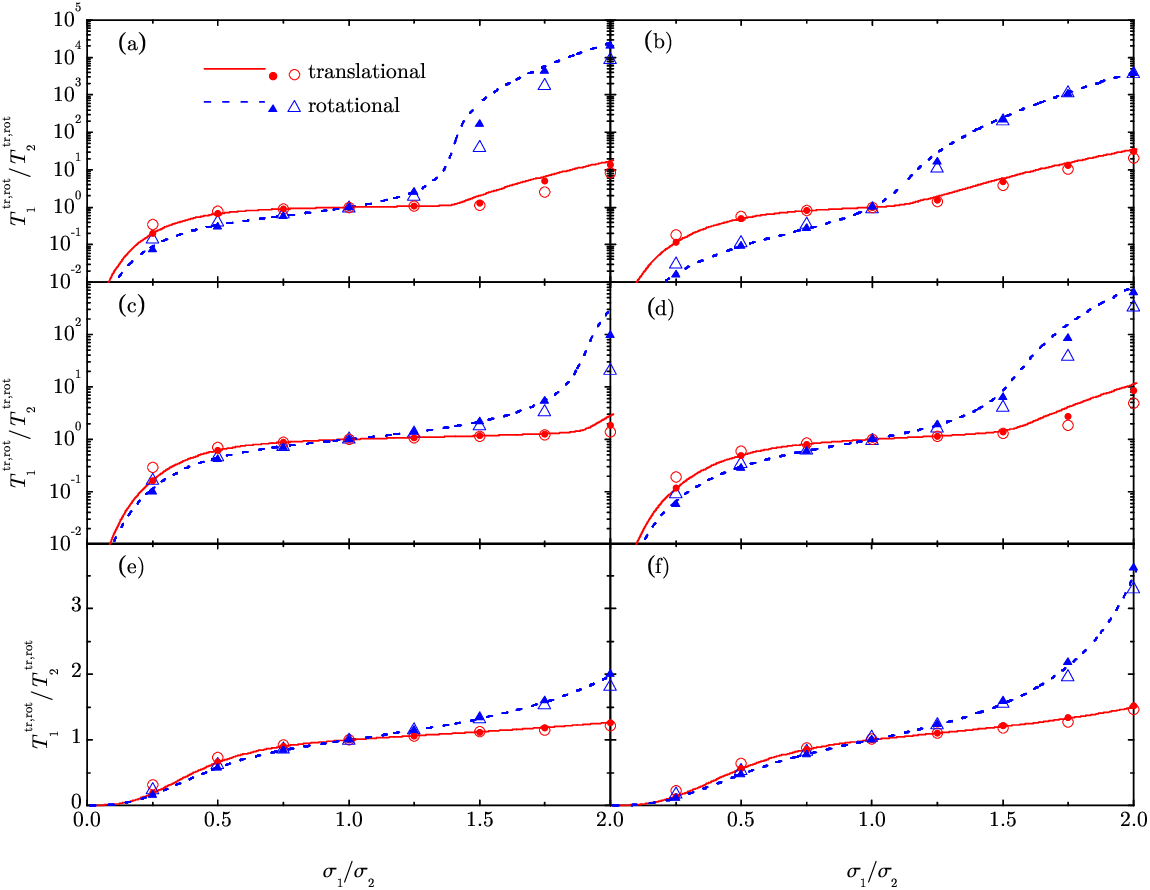}
\caption{Plot of the temperature ratios $T_1^\tr/T_2^\tr$ (solid lines and circles) and $T_1^\rot/T_2^\rot$ (dashed lines and triangles) versus $\sigma_1/\sigma_2$ for systems (a) A, (b) B, (c) C, (d) D, (e) E, and (f) F (see Table \ref{cases}). The lines are theoretical predictions, the filled symbols are DSMC results, and the open symbols are MD results. The error bars in the simulation data are smaller than the size of the symbols. Note the vertical logarithmic scales in panels (a)--(d).}
\label{fig:A-F}
\end{figure*}

\begin{figure*}
\includegraphics[width=1.8\columnwidth]{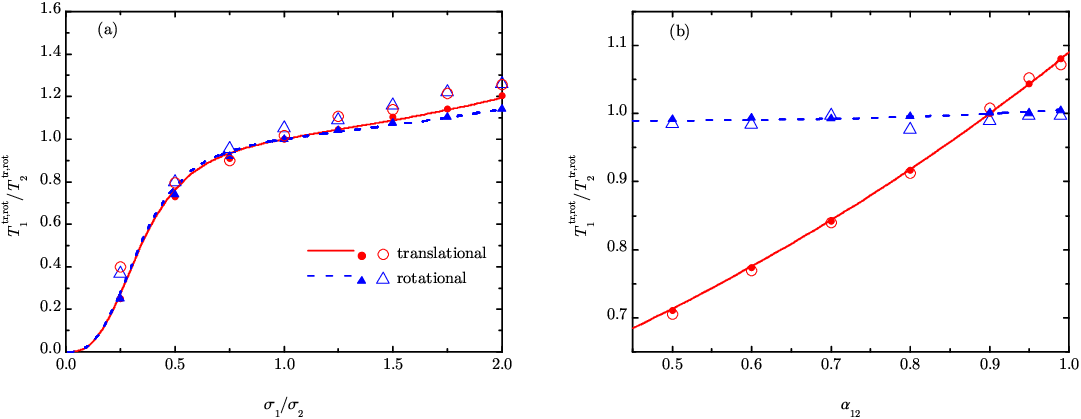}
\caption{Plot of the temperature ratios $T_1^\tr/T_2^\tr$ (solid lines and circles) and $T_1^\rot/T_2^\rot$ (dashed lines and triangles) versus (a) $\sigma_1/\sigma_2$  and (b) $\alpha_{12}$ for systems (a) G and (b) H (see Table \ref{cases}). The lines are theoretical predictions, the filled symbols are DSMC results, and the open symbols are MD results. The error bars in the simulation data are smaller than the size of the symbols.}
\label{fig:G-H}
\end{figure*}

\begin{figure*}
\includegraphics[width=1.8\columnwidth]{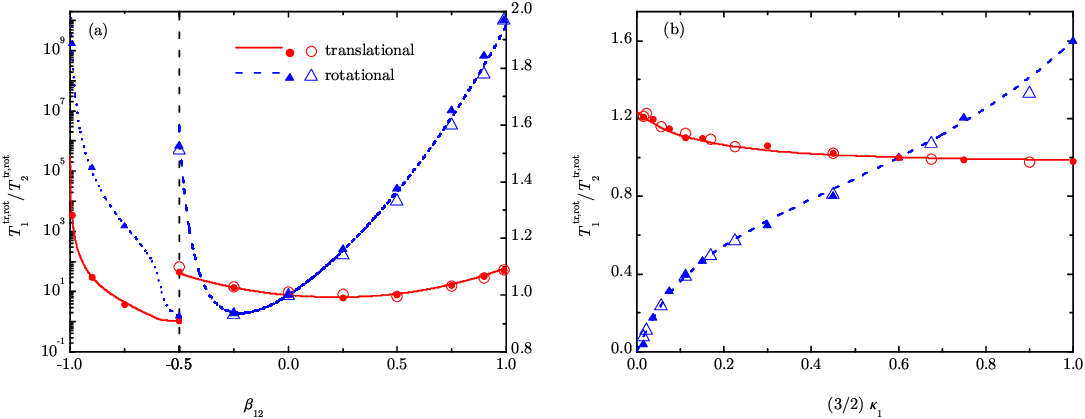}
\caption{Plot of the temperature ratios $T_1^\tr/T_2^\tr$ (solid lines and circles) and $T_1^\rot/T_2^\rot$ (dashed lines and triangles) versus (a) $\beta_{12}$  and (b) $\frac{3}{2}\kappa_1$ for systems (a) I and (b) J (see Table \ref{cases}). The lines are theoretical predictions, the filled symbols are DSMC results, and the open symbols are MD results. The error bars in the simulation data are smaller than the size of the symbols. Note that in panel (a) the vertical scale is logarithmic for $-1\leq\beta_{12}\leq -0.5$ and normal for $-0.5\leq\beta_{12}\leq 1$.}
\label{fig:I-J}
\end{figure*}

Before presenting the results for the partial temperatures associated with the tracer
component, it is worth  analyzing the temperature ratio $\Tr_2/\Tt_2$ for the host component.
According to the Maxwellian approximation, Eq.\ \eqref{IV.1}, the ratio  $\Tr_2/\Tt_2$ is a function of the parameter $\gamma_2$ only [see Eq.\ \eqref{Tr2}]. Figure \ref{fig:Tr2} shows the simulation values of $\Tr_2/\Tt_2$ as a function of $\gamma_2$ for the cases A--G. Notice that the host component in cases H--J is in the same state as in case C. We observe an excellent agreement with the Maxwellian prediction, although the latter tends to overestimate the ratio  $\Tr_2/\Tt_2$ for large values of the parameter $\gamma_2$.

Now we turn our attention to the tracer/host temperature ratios. Figure \ref{fig:A-F} shows the dependence of $\Tt_1/\Tt_2$ and $\Tr_1/\Tr_2$ on the size ratio $\sigma_1/\sigma_2$ for systems A-F.
From Fig.\ \ref{fig:A-F} it is quite apparent that the tracer temperature (either translational or rotational) is larger (smaller) than its host counterpart if the size of the tracer particle is larger (smaller) than that of a host particle. This trend is similar to what is observed in the case of smooth spheres \cite{MG02b,DHGD02}, where only translational temperatures play a role. This effect, however, is more pronounced in  $\Tr_1/\Tr_2$ than in $\Tt_1/\Tt_2$. Moreover, the disparity in temperatures increases as the roughness decreases. Comparison with the simulation data shows a good behavior of the approximate theoretical predictions.
An exception is the rotational temperature ratio $\Tr_1/\Tr_2$ at $\sigma_1/\sigma_2\geq 1.5$, where the MD data are typically smaller than both the theoretical and DSMC values. While this might be due in part to the non-negligible effect of a big and massive tracer particle on the properties of the host gas, it  also may reflect a breakdown of the molecular chaos assumption (at least in what concerns the distribution of angular velocities) in that situation.

In Fig.\ \ref{fig:G-H}(a) we display the results for system G, which corresponds to very inelastic spheres ($\alpha_{11}=\alpha_{22}=0.2$) with a  mass distribution concentrated near the center ($\kappa_1=\kappa_2=0.1$). The theory agrees quantitatively well with the DSMC data and semi-quantitatively with the MD simulations. This means that the molecular chaos hypothesis in the Boltzmann--Lorentz description is less reliable as the inelasticity increases, as expected from previous results \cite{DHGD02}. It is worthwhile noting that, in comparison with Fig.\ \ref{fig:A-F}, the breakdown of energy equipartition is much less significant in system G than in systems A--F.

In system H the impurity has the same mass, size, mass distribution, and coefficient of tangential restitution as the host particles, so that it only differs in the coefficient of normal restitution. This case has been analyzed by Goldhirsch and coworkers \cite{SGNT06,SNTG09} in the study of thermal diffusion segregation \cite{G11}. {}From Fig.\ \ref{fig:G-H}(b) we observe that the ratio of rotational temperatures has a very weak dependence on $\alpha_{12}$ and is quite close to $1$. This can be understood as a consequence of $\beta_{12}=\beta_{22}$. On the other hand, the different inelasticity has an obvious impact on the  ratio of translational temperatures. If $\alpha_{12}<\alpha_{22}$ the cooling effect is more significant for the tracer particle than for the host particles, which gives rise to $\Tt_1<\Tt_2$ in the asymptotic regime. The opposite happens when $\alpha_{12}>\alpha_{22}$. The agreement between theory and simulations is very good.

System I is analogous to system H, except that now  tracer and host particles have a different roughness. The results are displayed in Fig.\ \ref{fig:I-J}(a). The tracer/host temperature ratios reach very high values when the tracer particle is rather smooth. For instance, $\Tr_1/\Tr_2>10$ if $\beta_{12}\lesssim -0.58$ and $\Tt_1/\Tt_2>10$ if $\beta_{12}\lesssim -0.83$. For that reason, the vertical scale is logarithmic for $\beta_{12}<-0.5$ in Fig.\ \ref{fig:I-J}(a). In addition, in that region the duration of the transient regime toward the HCS increases considerably and, consequently, we were not able to reach steady values for the temperature ratios in our MD simulations. Interestingly, the temperature ratios have a nonmonotonic dependence on $\beta_{12}$ in the region $\beta_{12}>-0.5$. This behavior is very well captured by the approximate Maxwellian theory, as shown in Fig.\ \ref{fig:I-J}(a).

Finally, system J is studied in Fig.\ \ref{fig:I-J}(b). Now the only distinction between the tracer and host particles is the mass distribution, i.e., $\kappa_1\neq \kappa_2$. It is quite apparent that the impact of the mass distribution on the ratio of translational temperatures is rather weak, as might have been expected from the fact that the moment of inertia is directly related to the rotational degrees of freedom. In what concerns the ratio $\Tr_1/\Tr_2$, we observe that it grows as the mass of the tracer particle is more concentrated near the surface. Again, it  is worth highlighting  that all those features are quantitatively well accounted for by our Maxwellian approximation.

\section{Conclusions}
\label{sec5}
In this paper we have carried out a  study of the properties of an impurity (also
called tracer particle) immersed in a host gas of inelastic and rough hard spheres in the HCS regime. More specifically, we have focused on the quantities measuring the violation of energy equipartition between both species, namely the temperature ratios $\Tt_1/\Tt_2$ and $\Tr_1/\Tr_2$. Given the large number of independent parameters involved in the problem, we have considered $10$ different classes of systems (see Table \ref{cases}), in each one of them the tracer particle sharing the same mechanical properties as the host particles, except one. This has allowed us to offer a comprehensive view of the general tracer problem.

Three complementary but different routes have been followed. On the one hand, previous  results derived in Ref.\ \cite{SKG10} from the multitemperature Maxwellian approximation, Eq.\ \eqref{IV.1}, have been particularized to the problem at hand. All the quantities of the problem have been analytically determined in terms of the exact solution of an algebraic quartic equation. These theoretical results have been gauged via comparison with simulation results obtained independently from two methods: (i) a numerical solution of the Boltzmann and Boltzmann--Lorentz kinetic equations by the DSMC method and (ii) a MD simulation to solve Newton's equations of motion. The former  measures the accuracy of the Maxwellian approximation, while the latter assess the reliability of the molecular chaos assumption to compute the temperature ratios.

The results displayed in Figs.\ \ref{fig:A-F}--\ref{fig:I-J} show that, in general, the approximate theory compares well with simulations under conditions of practical interest. We have observed that limitations of the molecular chaos ansatz are noticeable only for large tracer particles and/or very inelastic gases, as shown in Figs.\ \ref{fig:A-F} and \ref{fig:G-H}(a). Even in those cases, the theoretical description agrees semi-quantitatively well with MD simulations.

The present study encourages us to address the more general problem of a binary mixture of inelastic and rough hard spheres. In that case, one has to deal with two coupled Boltzmann equations. In addition, the parameter space is augmented by three new parameters, namely the relative concentration  $n_1/n_2$ and the coefficients of restitution $\alpha_{11}$ and $\beta_{11}$. This will imply a much more stringent test for the multitemperature Maxwellian approximations. It will also be very interesting to consider, even in the tracer limit,  a moderately dense granular fluid and study the impact of velocity correlations on energy nonequipartition.
We plan to undertake those projects in the future.

\begin{acknowledgments}
This work has been supported by the
Ministerio de Econom\'ia y Competitividad (Spain) through Grant No.\
FIS2016-76359-P and  the Junta de Extremadura (Spain) through Grant
No.\ GR15104, both partially financed by ``Fondo Europeo de Desarrollo
Regional'' funds (FEDER). A.L.
acknowledges support from  Grant No.\
MTM2014-56948-C2-2-P and thanks the hospitality of the University of Extremadura, where  part of this work was done during a
stay. Use of computing facilities from the Extremadura Research
Centre for Advanced Technologies (CETA-CIEMAT), partially financed by the FEDER, is
also acknowledged.
\end{acknowledgments}

\appendix*
\section{Limit of a heavy impurity}
The general framework presented in Sec.\ \ref{sec3.B} is particularized here to  the case where the mass of a tracer particle is much larger than the mass of a host gas particle, i.e.\, $\mu\equiv m_1/m_2\gg 1$. In that limit [see Eqs.\ \eqref{20b}, \eqref{AA}, and \eqref{BB}],
\beq
\kappa_{12}\approx \kappa_2,\quad A\approx\mu^{-1}(1+\alpha_{12}),\quad  B\approx \frac{\kappa_{2}}{1+\kappa_{2}}\frac{1+\beta_{12}}{1+\alpha_{12}}.
\eeq

Three independent possibilities must be distinguished.

\subsection{$\displaystyle{1+\alpha_{12}>\frac{\kappa_2(1-\kappa_1)}{\kappa_1(1+\kappa_2)}(1+\beta_{12})}$}
\label{subsecA1}

By assuming that $\mu Y\gg X\gg 1$, Eq.\ \eqref{56.1} yields
\begin{subequations}
\label{X+Y}
\beq
X\approx \left[\frac{\zeta^*\kappa_1}{2sB(1+\alpha_{12})}\right]^2{\mu^2}.
\label{X}
\eeq
Next, we consider Eq.\ \eqref{55.1}. If one had $X\gg \mu^{-1}Y$, then Eq.\ \eqref{55.1} would provide an expression for $X$ inconsistent with \eqref{X}. Thus, we conclude that $Y\sim \mu X$, so that
\beq
Y\approx 2\frac{1-B(\kappa_1^{-1}-1)}{B^2 r(1+\alpha_{12}) }\mu {X}.
\label{Y}
\eeq
\end{subequations}
Going back to the temperature ratios [see Eq.\ \eqref{XY}],
\begin{subequations}
\label{2.2}
\bal
\frac{T_1^\tr}{T_2^\tr}\approx& C_{\text{I}}^2\left(\frac{m_1}{m_2}\right)^3
\left(\frac{\sigma_{2}}{\sigma_{12}}\right)^4,
\label{2.2.1}
\\
\frac{T_1^\rot}{T_1^\tr}\approx & K_{\text{I}}\frac{m_1}{m_2},
\eal
\end{subequations}
where
\begin{subequations}
\label{2.2b}
\bal
\label{CI}
C_{\text{I}}\equiv& \frac{3\zeta^*\kappa_1(1+\kappa_2)\chi_{22}/\chi_{12}}{4\sqrt{2}\kappa_2(1+\beta_{12})},
\\
\label{KI}
K_{\text{I}} \equiv& \frac{2\kappa_1(1+\kappa_2)^2}{\kappa_2^2}\frac{1+\alpha_{12}-\frac{\kappa_2(1-\kappa_1)}{\kappa_1(1+\kappa_2)}(1+\beta_{12})}{(1+\beta_{12})^2}.
\eal
\end{subequations}
Therefore, the asymptotic behaviors are
\begin{subequations}
\label{asymptA}
\bal
\frac{T_1^\tr}{T_2^\tr}\sim &\left(\frac{m_1}{m_2}\right)^3\left(\frac{\sigma_{2}}{\sigma_{12}}\right)^4,
\\
\frac{T_1^\rot}{T_2^\rot}\sim &\left(\frac{m_1}{m_2}\right)^4\left(\frac{\sigma_{2}}{\sigma_{12}}\right)^4.
\eal
\end{subequations}
If the impurity and bath particles have comparable mass densities, then $m_1/m_2\sim (\sigma_1/\sigma_2)^3$, so that Eqs.\ \eqref{asymptA} imply (by ignoring the dependence of $\chi_{22}/\chi_{12}$ on the size ratio) that $T_1^\tr/T_2^\tr\sim (\sigma_1/\sigma_2)^5$ and $T_1^\rot/T_2^\rot\sim (\sigma_1/\sigma_2)^8$ in the limit $\sigma_1/\sigma_2\gg 1$.

It must be noted that the above analysis (based on the assumptions $Y\sim \mu X\gg X\gg 1$) is valid inasmuch as the predicted temperature ratios are positive definite, i.e., if, according to Eq.\ \eqref{Y}, $B(\kappa_1^{-1}-1)< 1$. This is equivalent to [see Eq.\ \eqref{KI}]
\beq
1+\alpha_{12}>\frac{\kappa_2(1-\kappa_1)}{\kappa_1(1+\kappa_2)}(1+\beta_{12}).
\label{2.1}
\eeq
On the other hand, a Monte Carlo estimate with $10^8$ sets of random values $0\leq\alpha_{12}\leq 1$, $-1\leq\beta_{12}\leq 1$, $0\leq\kappa_{2}\leq \frac{2}{3}$,  and $0\leq\kappa_{1}\leq \frac{2}{3}$ shows that Eq.\ \eqref{2.1} is violated about $19.26\%$ of all the possible cases.
For instance, if $\kappa_2=\frac{2}{5}$, $\alpha_{12}=\frac{1}{2}$, and $\beta_{12}=1$, Eq.\ \eqref{2.1} is not fulfilled if $\kappa_1\leq \frac{8}{29}\simeq 0.276$.

Obviously, when Eq.\ \eqref{2.1} is not  verified, Eqs.\ \eqref{X+Y}--\eqref{asymptA} are no longer applicable. In that case,  the working hypothesis $\mu Y\gg X\gg 1$  must be replaced by $\mu Y\sim X\gg 1$.

\subsection{$\displaystyle{1+\alpha_{12}<\frac{\kappa_2(1-\kappa_1)}{\kappa_1(1+\kappa_2)}(1+\beta_{12})}$}
\label{subsecA2}

By assuming $\mu Y\sim X\gg Y\gg 1$, Eqs. \eqref{55.1} and \eqref{56.1} yield, respectively,
\begin{subequations}
\beq
X\approx \left[\frac{\zeta^*}{2s(1+B)(1+\alpha_{12})}\right]^2{\mu^2},
\label{X2}
\eeq
\beq
Y\approx \frac{B^2(1+\alpha_{12})/2r\kappa_1^{2}}{B(\kappa_1^{-1}-1)-1}\frac{X}{\mu}.
\label{2.4}
\eeq
\end{subequations}
Going back to the original quantities,
\begin{subequations}
\label{2.3}
\bal
\frac{T_1^\tr}{T_2^\tr}\approx& C_{\text{II}}^2\left(\frac{m_1}{m_2}\right)^3
\left(\frac{\sigma_{2}}{\sigma_{12}}\right)^4,
\label{2.3.1}
\\
\frac{T_1^\rot}{T_1^\tr}\approx & K_{\text{II}}\frac{m_2}{m_1},
\eal
\end{subequations}
with
\begin{subequations}
\label{2.3b}
\bal
\label{CII}
C_{\text{II}}\equiv& \frac{3\zeta^*\chi_{22}/\chi_{12}}{4\sqrt{2}\left[1+\alpha_{12}+\frac{\kappa_2}{1+\kappa_2}(1+\beta_{12})\right]},
\\
\label{KII}
K_{\text{II}}\equiv& -\frac{1}{K_{\text{I}}}.
\eal
\end{subequations}
The asymptotic behaviors are of the form
\begin{subequations}
\label{asymptB}
\bal
\frac{T_1^\tr}{T_2^\tr}\sim &\left(\frac{m_1}{m_2}\right)^3\left(\frac{\sigma_{2}}{\sigma_{12}}\right)^4,
\\
\frac{T_1^\rot}{T_2^\rot}\sim &\left(\frac{m_1}{m_2}\right)^2\left(\frac{\sigma_{2}}{\sigma_{12}}\right)^4.
\eal
\end{subequations}
Now, if $m_1/m_2\sim (\sigma_1/\sigma_2)^3$, then  $T_1^\tr/T_2^\tr\sim (\sigma_1/\sigma_2)^5$ and $T_1^\rot/T_2^\rot\sim (\sigma_1/\sigma_2)^2$ in the limit $\sigma_1/\sigma_2\gg 1$.

\subsection{$\displaystyle{1+\alpha_{12}=\frac{\kappa_2(1-\kappa_1)}{\kappa_1(1+\kappa_2)}(1+\beta_{12})}$}

This is the threshold between the cases \ref{subsecA1} and \ref{subsecA2} above. It takes place if $B(\kappa_1^{-1}-1)=1$, i.e., if both sides of Eq.\ \eqref{2.1} are equal. Now $X$ and $Y$ are of the same order, $Y\sim X\gg 1$.
Because of the same reasoning as in case \ref{subsecA1}, Eqs.\ \eqref{X} and \eqref{2.2.1} still apply. In fact, in this special case they are equivalent to Eqs.\ \eqref{X2} and \eqref{2.3.1}, respectively, i.e., $C_{\text{I}}=C_{\text{II}}$.
Next, by equating the right-hand sides of Eqs.\ \eqref{55.1} and \eqref{56.1} and using $B=\kappa_1/(1-\kappa_1)$, one finds a quadratic equation
for ${\kappa_1 r Y}/{X}\equiv {T_1^\rot}/{T_1^\tr}$,
\beq
\left(\frac{T_1^\rot}{T_1^\tr}\right)^2-2(1-\kappa_1)\frac{T_1^\rot}{T_1^\tr}-1=0,
\label{2.5}
\eeq
whose positive solution is
\beq
\frac{T_1^\rot}{T_1^\tr}\approx K_{\text{III}},\quad
K_{\text{III}}\equiv   1-\kappa_1+\sqrt{(1-\kappa_1)^2+1}.
\label{2.6}
\eeq

Table \ref{tablelimit} summarizes the results of this Appendix.

%\bibliographystyle{apsrev}
%\bibliography{D:/Dropbox/Mis_Dropcumentos/bib_files/Granular}

\end{document}